# Burst Transmission Symbol Synchronization in the Presence of Cylce Slip Arising from Different Clock Frequencies

Somaye Bazin

bazin.somayeh@gmail.com

Mahmoud Ferdosizade Naeiny

Electrical Engineering Department, Shahed Univeristy

Ferdosizadeh@shahed.ac.ir

Roya Khanzade

Electrical Engineering Department, Shahed Univeristy

royakhanzade@gmail.com

*Abstract*—In digital communication systems different clock frequencies of transmitter and receiver usually is translated into cycle slips. Receivers might experience different sampling frequencies from transmitter due to manufacturing imperfection, Doppler Effect introduced by channel or wrong estimation of symbol rate. Timing synchronization in presence of cycle slip for a burst sequence of received information, leads to severe degradation in system's performance that represents as shortening or prolonging of bit stream. Therefor the necessity of prior detection and elimination of cycle slip is unavoidable. Accordingly, the main idea introduced in this paper is to employ the Gardner Detector (GAD) not only to recover a fixed timing offset, its output is also processed in a way such that timing drifts can be estimated and corrected. Deriving a two steps algorithm, eliminates the cycle slips arising from wrong estimation of symbol rate firstly, and then iteratively synchronize symbol's timing of a burst received signal by applying GAD to a feed forward structure with the additional benefits that convergence and stability problems are avoided, as they are typical for feedback schemes normally used by GAD. The proposed algorithm is able to compensate considerable symbol rate offsets at the receiver side. Considerable results in terms of BER confirm the algorithm's proficiency.

1. INTRODUCTION

Timing recovery as a process of sampling at the right times is critical in digital communication receivers. In principle, the problem is formulated through Maximum Likelihood (ML). The direct computation of ML problem, performs the task of jointly detection and estimation of the message bits and the timing offset. However this solution conveys the exhaustive search methods that imposes a lot of computation and makes the solution impractical. To avoid the complexity of direct computation of ML problem, iterative solutions are introduced [1-3]. The general idea of iterative timing recovery scheme is to improve the timing estimation accuracy by multiple exploiting the timing information provided by a set of samples and application of this estimation to regenerate a new set of samples that iteratively approaches to the local maximum of the likelihood function. ML based timing recovery methods usually ignore the time varying timing offsets and proceed under the assumption of fixed synchronization parameter estimation [4-6]. However in practice, the timing offset may vary with time, due to the different clock frequencies in transmitter and receiver caused by fractional error in baud rate estimation, manufacturing imperfection and etc. [7].

Different clock frequencies in transmitter and receiver leads to linear increasing of timing offset from symbol to symbol. While, timing offset increases linearly for successive symbols, Synchronizers may fail to track this time varying delay. Since getting far away from true value makes the estimator falls into the adjacent stable operating point and synchronizer starts to keep tracking this new stable operating point. Consequently one symbol inserted into or erased from the sequence. This is called Cycle Slipping (CS). There is also another source of CS which is the large phase variance of voltage controlled oscillators (VCOs) caused by low signal to noise ratio (SNR) that is not subjected in this paper. As long as cycle slips occur, system's performance decreases dramatically due to the relative loss of synchronization caused by symbol insertion or omission in the sequence. In order to alleviate the adverse effect of cycle slip it has to be eliminated before applying timing synchronization.

Although, several studies have considered the problem of cycle slipping in synchronizers [8-10], few authors have proposed the solution [11-13]. Typically, solutions concern about error tracking synchronizers in low SNRs which are based on closed feedback loop. While the good tracking performance of feedback schemes is not deniable, they require, in counterpart relatively long acquisition time that makes them unsuitable for burst transmission schemes .In this sense, a feed forward structure based on extracting timing delay estimation from the statistics of received samples, and then adjusting the time by some sort of interpolation is more suitable. In this work, in order to utilize the band width efficiency of Non-Data-Aided (NDA) estimators and effective flexibility of interpolation, Gardner TED [14] and Farrow filter [15] are used in a feed forward structure as timing delay estimator and interpolation filter respectively.

In accordance with the above statements, this work is motivated by the objective of deriving a novel algorithm which employs GAD in a non-conventional manner so that not only the fixed timing offset is recovered, GAD's output is also processed in a way such that considerable timing drifts can be estimated and corrected, which is not addressed in the literatures. CS is eliminated at the first step and then Gardner timing delay estimation in cooperation with Farrow based interpolation filter Makes the algorithm suitable for a more agile and efficient iterative receiver.

2. PROBLEM FORMULATION

2.1 *Signal model*

Assume a traditional communication system, where the transmitted signal is corrupted by passing through AWGN channel which also imposes a timing delay and carrier frequency and phase offset to the received signal as follows:

$$r(t) = e^{j2\pi\Delta f t + \theta} \sum_{n=0}^{N-1} a_n h(t - nT - \tau) + n(t) \tag{1}$$

Where $a_n$ denotes the zero mean unit variance, independently and identically distributed (i.i.d.) symbols that might be taken from any linear modulation scheme. $\theta$ and $\Delta f$ are phase and carrier frequency offsets. $h(t)$ is a square root raised cosine pulse and $n(t)$ is a complex zero-mean additive white Gaussian noise with two sided power spectral density of $N_0/2$. Moreover $\tau$, $T$ and $N$ are unknown timing delay, pulse duration and the length of the transmitted signal respectively.

At the receiver side, the signal in (1) should be matched filtered and the transmitted symbols should be regenerated by sampling $r(t)$ in $kT - \hat{\tau}$ time instants, where $\hat{\tau}$ is timing delay estimation provided by synchronizer. Even if receiver has exact information about symbol rate, there still may exist some fractional difference between transmitter and receiver clock frequencies due to the implementation imperfection. However, blind receivers have to estimate symbol rate that always conveys some estimation error. In this case the regenerated symbols are located in $kT' - \hat{\tau}$ time instants, where:

$$T' = T + \varepsilon \tag{2}$$

and $\varepsilon$ is the difference between transmitter and receiver symbol duration and can be either positive value or negative one. The discrete samples are:

$$r(k) = e^{j(2\pi\Delta f t + \theta)} \sum_{n=0}^{N-1} (a_n + w_n) g(kT' - \hat{\tau} - nT + \tau)$$

$$= e^{j(2\pi\Delta f t + \theta)} \sum_{n=0}^{N-1} A_n g(kT - nT + (\mu + k\varepsilon)) \tag{3}$$

Where $w_n$ is i.i.d. normally distributed variable with variance $\sigma^2$, $g(t)$ is convolution of $h(t)$ with channel impulse response and the analog pre-filter response. $A_n$ represents $a_n + w_n$ and $\mu$ stands for $\tau - \hat{\tau}$.

Obviously, timing delay varies for different symbols of a received burst, due to the variable timing delay part which is increased linearly by $k$. Traditional approaches assume this variation is slow in comparison with symbol interval and they approximate the timing delay over a number of symbol periods that synchronization parameter can be considered as quasi-constant [16], however ignoring this variation would degrade the performance as it will be well illustrated in simulation results.

### 2.2 Cycle-Slips in synchronizer

Typically, Gardner's Timing Error Detection (TED) provides timing estimation to synchronize the received symbols using the samples at twice rate of the symbol rate, according to the following equation [14]:

$$u(k) = \Re\{r^*(k+1/2)[r(k+1) - r(k)]\} \tag{4}$$

Here $u(k)$ is the timing error of the kth symbol, $\Re\{.\}$ is the real part and * denotes complex conjugation. Plugging (3) into (4) results:

$$u(k) = \Re\{e^{-j(2\pi\Delta f(k+\frac{1}{2})T+\theta)} \sum_{n=0}^{N-1} A_n^* g^*\left(\left(k+\frac{1}{2}\right)T - nT + \left(\mu + \left(k+\frac{1}{2}\right)\varepsilon\right)\right)$$
$$* [e^{j(2\pi\Delta f(k+1)T+\theta)} \sum_{n'=0}^{N-1} A_{n'} g((k+1)T - n'T + (\mu + (k+1)\varepsilon)) - e^{j(2\pi\Delta f kT+\theta)} \sum_{n''=0}^{N-1} A_{n''} g(kT - n''T + (\mu + k\varepsilon))]\} \tag{5}$$

The obvious fact is that the phase offset does not play an influential role in Gardner's timing delay estimation, likewise the impression of carrier frequency offset on timing delay estimation is negligible, considering an ordinary assumption that $\Delta fT \ll 1$. Therefore carrier and phase offset can be omitted as far as $\frac{\Delta f}{BW} \ll 1$ and timing offset can be synchronized regardless of prior carrier and phase synchronization.

As long as $\mu + k\varepsilon < |T/2|$ TED is capable of tracking the timing delay. For any special value of $k$ that $\mu + k\varepsilon$ exceeds from this interval the synchronizer starts to relate the timing delay to the adjacent symbols and CS happens. This non-uniform delay detection results in a quasi-periodic function of $u(k)$ which is wrapped for particular coefficients of $K$ that $K\varepsilon = T$. The periodicity of $u(k)$ is proved in the following section.

### 2.2.1 Lemma

For the sake of simplicity, transmitted signal supposed to be a burst sequence that alternatively changes between -1 and +1 and $g(t)$ is band limited with no excess double sided bandwidth of $1/T$, as it is depicted in figure 1. Also delay offset for three successive samples involving for symbol's offset computation in (5) is assumed to be same value equals $\mu + k\varepsilon$.

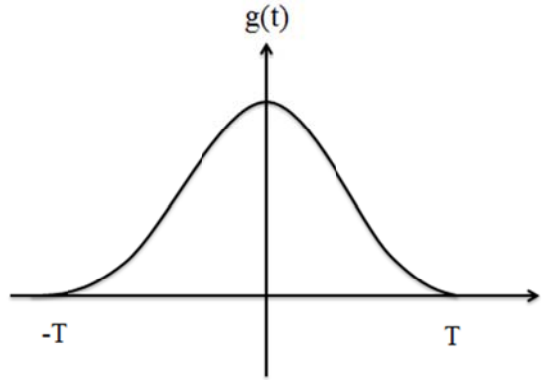

Figure .1 g(t) supposed to be bandlimited.

Considering bandwidth of $g(t)$ and doing some manipulation in order to discard ineffective sentences in (5), a simplified extend is obtained, As long as $|\mu + k\varepsilon| \le \frac{T}{2}$ Let $u_1(k)$ represents the timing offset estimation of $k$th symbol. For any arbitrary positive value of $\delta$ that $0 < \delta < \frac{T}{2}$ and $\mu + k\varepsilon = \frac{T}{2} - \delta$, $u_1(k)$ is generated as follows:

$$u_1(k) = [A_k{}^* g^*(T-\delta) + A_{k+1}{}^* g^*(-\delta)] \times$$
$$[(A_{k+1} - A_K)g\left(\frac{T}{2} - \delta\right) + (A_{k+2} - A_{k+1})g\left(-\frac{T}{2} - \delta\right)] \quad (6)$$

Or:

$$u_1(k) = [A_k{}^* g^*(T-\delta) + A_{k+1}{}^* g^*(-\delta)]$$
$$\times [A_{k+1}\left(g\left(\frac{T}{2} - \delta\right) - g\left(-\frac{T}{2} - \delta\right)\right)$$
$$+ \left(A_{k+2} g\left(-\frac{T}{2} - \delta\right) - A_k g\left(\frac{T}{2} - \delta\right)\right)] \quad (7)$$

Evidently:

$$g(T-\delta) \le g(\delta)$$
$$g\left(-\frac{T}{2} - \delta\right) \le g\left(\frac{T}{2} - \delta\right)$$

Noting to the fact that $A_{k+2}$ and $A_k$ are even +1 or -1, term$\left(A_{k+2} g\left(-\frac{T}{2} - \delta\right) - A_k g\left(\frac{T}{2} - \delta\right)\right)$ can be ignored in (7) and consequently:

$$sign\left(u_1(k)\right) = |sign(A_{k+1})|^2$$

Which is always positive irrespective to what $A_k$ and $A_{k+1}$ are, and confirm the correctness of delay estimation.

Suppose that for all $k$th symbols that $\mu + k\varepsilon > |\frac{T}{2}|$, $u_2(k)$ represents the relevant timing offset estimation, Similarly assume $\delta$ is a positive value, $0 < \delta < \frac{T}{2}$, that make the delay exceeds from $\frac{T}{2}$, so that $\mu + k\varepsilon = \frac{T}{2} + \delta$. Relatively $u_2(k)$ can be achieved through following equation:

$$u_2(k) = [A_{k+1}{}^* g^*(\delta) + A_{k+2}{}^* g^*(\delta - T)] \times$$
$$[(A_{k+1} - A_k)g\left(\delta + \frac{T}{2}\right) + (A_{k+2} - A_{k+1})g\left(\delta - \frac{T}{2}\right)] \quad (8)$$

Regenerating of $u_1(k)$ for $\mu + k\varepsilon = -\frac{T}{2} + \delta$ corresponds to:

$$u_1(k) = [A_k{}^* g^*(\delta) + A_{k+1}{}^* g^*(\delta - T)]$$
$$[(A_k - A_{k-1})g\left(\delta + \frac{T}{2}\right) + (A_{k+1} - A_k)g\left(\delta - \frac{T}{2}\right)] \quad (9)$$

It is concluded from (8) and (9) that:

$$u_1(k+1) = u_2(k) \tag{10}$$

Or equivalently as long as timing delay increasing, caused by $k\varepsilon$, varies in $\left[-(T/2-\delta):\left(\frac{T}{2}-\delta\right)\right)$, synchronizer is able to track and detect the relative delay. Once timing delay exceeds from this span, TED periodically will assign the error to the adjacent symbol, so that, delay greater than $T/2$ would not be estimated and consequently CS happens alternatively. This is shown in figure.2. This non-uniform delay detection results in a quasi-periodic function of $u(k)$ which is wrapped for particular coefficients of $K$ that $K\varepsilon = T$. Conventionally, variable timing delay offset is assumed to be negligible by restricting the number of $k$ so that $k \ll K$. However, this contribution is concerned about the problem of at least one cycle slip takes place during the received burst sequence which is the case of either long burst with small $\varepsilon$ or short burst with significant $\varepsilon$.

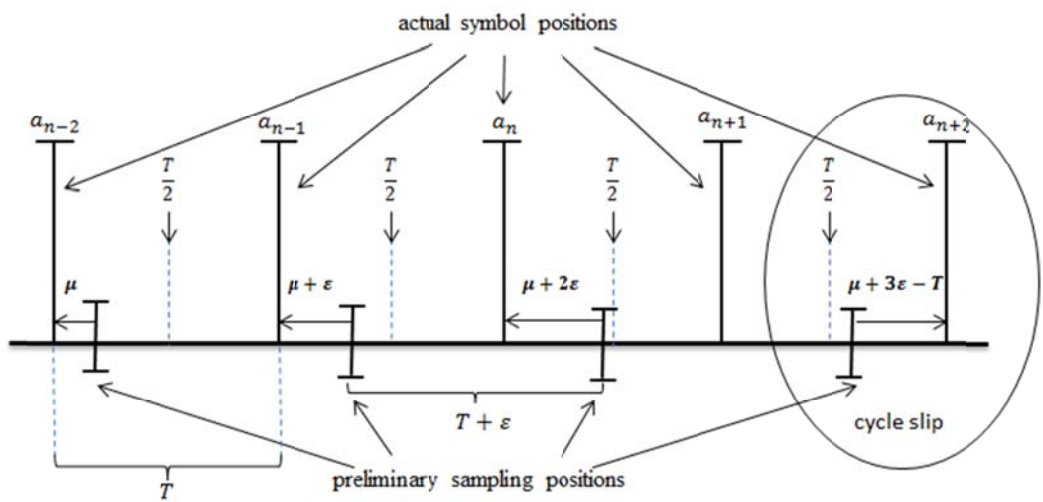

Figure.2. illustration of cycle slip in synchronizer

2.3 *Cycle-Slips detection and correction*

Let $R_s$ and $\hat{R}_s$ denote the symbol rate of the transmitted signal and respective estimation of symbol rate at the receiver, where:

$$\hat{R}_s = R_s + \epsilon R_s \tag{8}$$

and $\epsilon \in [-1, 1]$ is the normalized symbol rate offset. Equivalently the relation between $T$ and $T'$ can be expressed as:

$$T' = T - \frac{\epsilon}{1+\epsilon} T \tag{9}$$

Suppose $K$ is a maximum number of symbols that synchronizer can track without cycle slipping. By a little manipulation the following equation can be obtained:

$$K = \frac{1+\epsilon}{\epsilon} \quad (10)$$

$K$ Can be interpreted as a period of $u(k)$. Fig.1 represents a realization of $u(k)$, plotted for a received burst signal with the length of 500 symbols, where $\epsilon = 0.1$. As it is demonstrated although the mathematical expression of alternate cycle slip is straight forward, $u(k)$ is a noisy version of a periodic signal, This is mainly due to the Gardner's TED self-noise arising from data randomness [17] and the additive noise impairment of channel. Once the periodic term of $u(k)$ is extracted, it can be utilized in order to detect and correct cycle slip.

CS existence is determined by using Discrete Fourier Transform (DFT). Suppose $\mathbf{u} = [u(1), u(2), \ldots u(k), \ldots u(N)]$ and $\mathbf{U} = [U(1), U(2), \ldots U(l), \ldots U(L)]$, where $\mathbf{U} = \mathcal{F}\{\mathbf{u}\}$. And $\mathcal{F}\{\}$ stands for Discrete Fourier Transform and defined as $U(l) = \sum_{k=0}^{N-1} u(k) e^{j2\pi kl/L}$. Let:

$$q = argmax_{l=0,\ldots,L-1} U(l)$$

Fig.2 shows the DFT of depicted signal in fig.1. Clearly, the periodic term of $u(k)$ which is an indication of CS happening, results in a dominant component in the frequency of $1/K$. While constant timing delay without CS, yields a prominent DC component in DFT of $u(k)$. In other words:

$$q = 0 \quad \text{no cycle slip}$$
$$q \neq 0 \quad \text{cycle slip}$$

However, in some cases it might be more practical to compare $q$ with small threshold instead of absolute zero due to the noise existence. Obviously, when CS is recognized, DFT can be used to extract $K$ by:

$$K = \frac{L}{q} \quad (11)$$

In (11) $K$ is approximated by $L$-point DFT where, $L > N$ and $N$ is the length of $u(k)$. It is equivalent to first zero padding of $u(k)$ and then application of DFT for efficient implementation, and finally the index that maximize the DFT results in an approximation of $K$.

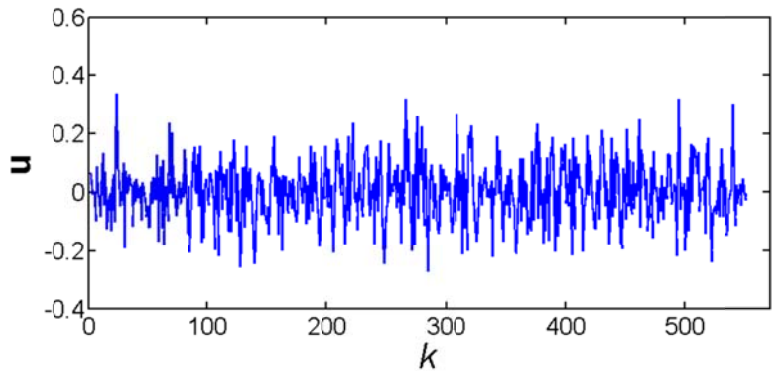

Figure.1. **u** for a burst length of 500 symbol, modulation type: BPSK, $\epsilon = 0.1$, SNR = 10db. CS leads to burst of error occurs due to the symbol insertion.

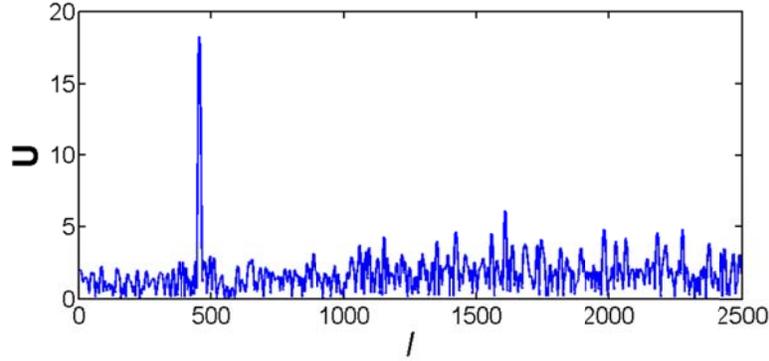

Figure.2. 5000- point DFT of $u$ Burst length: 500 symbol, modulation type: BPSK, $\epsilon = 0.1$, SNR = 10db. CS is represented as a frequency in 5000/K.

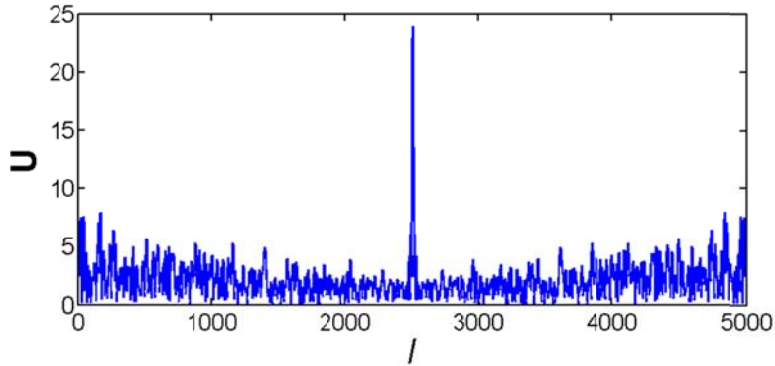

Figure.3. 5000- point DFT of $u(k)$ after CS correction. Burst length: 500 symbol, modulation type: BPSK, SNR = 10db. Constant timing delay is represented as DC component.

After determination of $K$, using (8) and (10), CS would be eliminated by improving the estimation of symbol rate in the receiver, according to the following equation:

$$R_s = (1 - \frac{1}{K}) \hat{R}_s \qquad (12)$$

Modification of the symbol rate estimation, leads to CS correction and the periodic term removal in **u**. Once the $R_s$ correction is done the CS detection procedure repeats again to make sure of periodic term elimination and if any periodic term exists, CS correction process repeats. However, CS elimination is not sufficient for timing recovery, since it is not able to recover the constant timing delay $\mu$, that translates itself into a DC component after CS elimination. Fig.3 represents DFT of $u(k)$ after CS correction and the general block diagram of proposed algorithm is depicted in figure 4.

## 3 ITERATIVE TIMING RECOVERY

In [18], an iterative timing recovery derivation from the maximum likelihood principle is proposed, where the timing information extracted from Gardner's TED is used iteratively to adjust sampling time. Here this method is used in cooperation with digital filtering by Farrow interpolation structure in order to improve the sampling instants.

Suppose $U(0)$ is the DC component of **U** clearly:

$$U(0) = \sum_{k=0}^{N-1} u(k) \tag{13}$$

And $\mu$ would be approximated as:

$$\mu = \frac{1}{N} U(0) \tag{14}$$

In the other words, averaging the evaluated timing error by Gardner's TED over all the symbols of a burst provides an estimation of constant timing delay. This can be used for timing adjustment in interpolator.

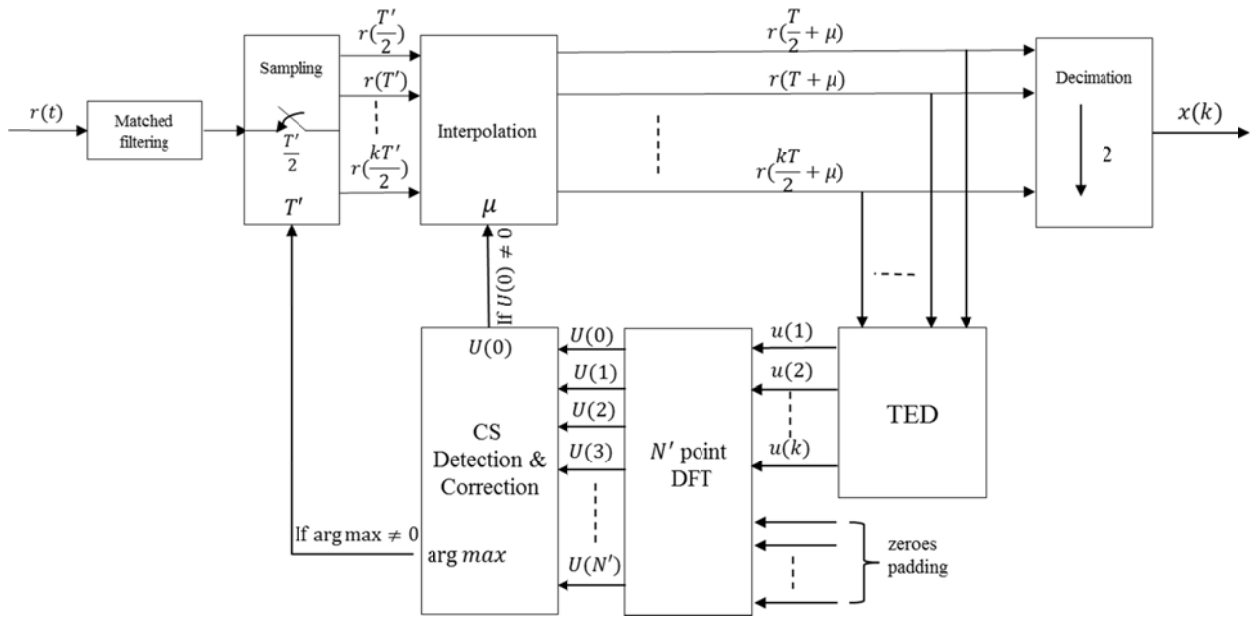

Figure.4. System block diagram

In those cases that the fractional delay is sufficient to control the interpolator's structure and there is no computational need to update the filter's coefficients, Farrow structure of polynomial interpolation is preferred due to the implementation efficiency. The detailed explanation of Farrow structure can be find in [19]. While different polynomial orders can be considered in order to generate the outputs, applying Lagrange polynomial in Farrow structure when the polynomial order, for example, corresponds to 3 result in a following equation for interpolation's output.

$$x(k) = \mu^3 \left(-\frac{1}{6}r_{-1} + \frac{1}{2}r_0 - \frac{1}{2}r_1 + \frac{1}{6}r_2\right)$$
$$+ \mu^2 \left(-\frac{1}{2}r_{-1} + 2r_0 - \frac{5}{2}r_1 + r_2\right) \qquad (15)$$
$$+ \mu \left(-\frac{1}{3}r_{-1} + \frac{3}{2}r_0 - 3r_1 + \frac{11}{6}r_2\right) + r_2$$

Figure. 5 illustrates the respective samples involving to generate the interpolation's output and $\mu$. Note that in order to exploit the timing information by Gardner's TED, two samples are generated for each symbol timing interval. Once the new samples are obtained by interpolation, they are used to evaluate $u(k)$, then applying (13) and (14) results in a new approximation of $\mu$. If $\mu \cong 0$, the generated samples are decimated by the factor of 2 and the selected samples are used to determine the final symbols, otherwise the timing delay is modified using (18) and this procedure is repeated. Iterative application of this algorithm leads to the timing recovery and it locates the samples at the right timing positions.

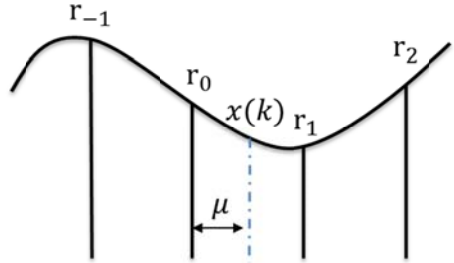

Figure.5 illustration of interpolation

4 SIMULATION RESULTS

To verify the algorithm's proficiency, simulation results are presented in this section. Simulations are carried out with BPSK and QPSK signals which are shaped by a square root raised cosine filter with roll-off factor of 0.5 in transmitter and they are passed through AWGN channel. The interpolation filter order in the timing recovery block is set to 3 and the DFT length for CS correction is 5000.

In fig 6 system's performance is evaluated and compared in terms of BER of the BPSK signal modulation. The proposed algorithm is referred as "CS corrected timing recovery". While timing recovery without CS correction is referred as "CS uncorrected timing recovery". The burst length of the received signal is 300 symbols, and the normalized symbol rate offset $\epsilon$ is set to 0.1. As it is shown, despite the significant decrease in performance, due to the CS existence in CS uncorrected timing recovery plot, BER has improved dramatically by prior elimination of CS through introduced algorithm in CS corrected timing recovery plot. In fig 7, "burst by burst timing recovery" imply to the typical approach that it synchronizes the timing delay over the number of Symbols' interval in which, timing delay can be assumed piecewise constant and the variation can be ignored.

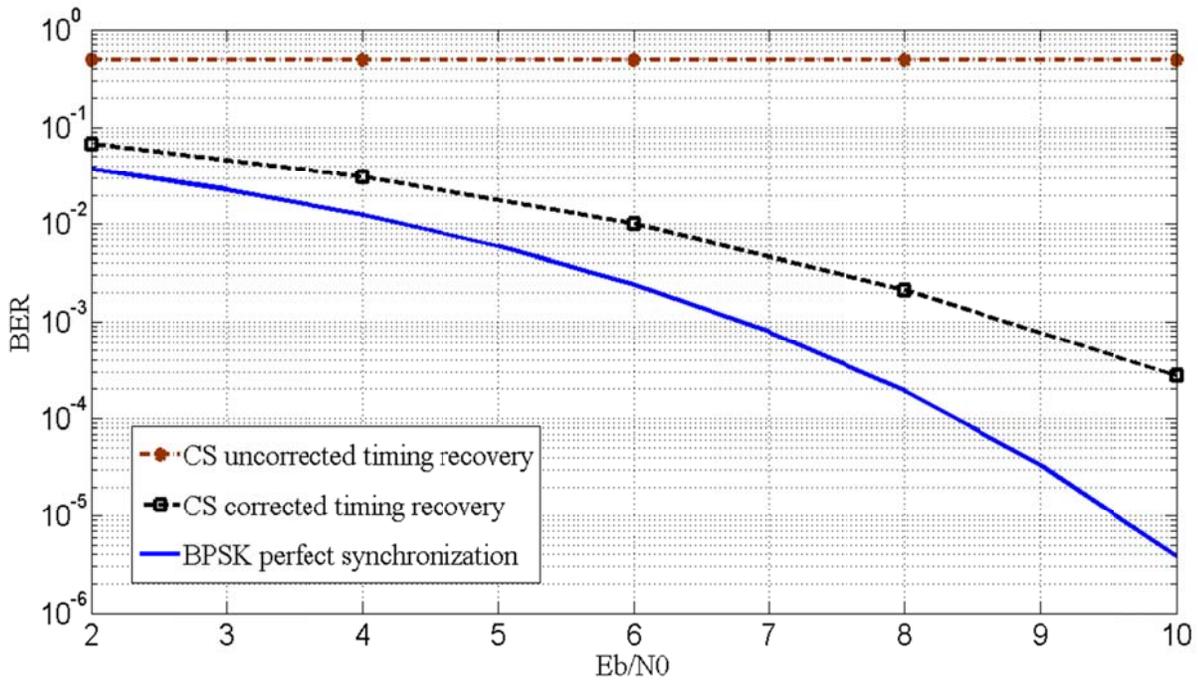

Fig. 6. BER performance, $N = 300$, $\epsilon = 0.1$, $L = 5000$

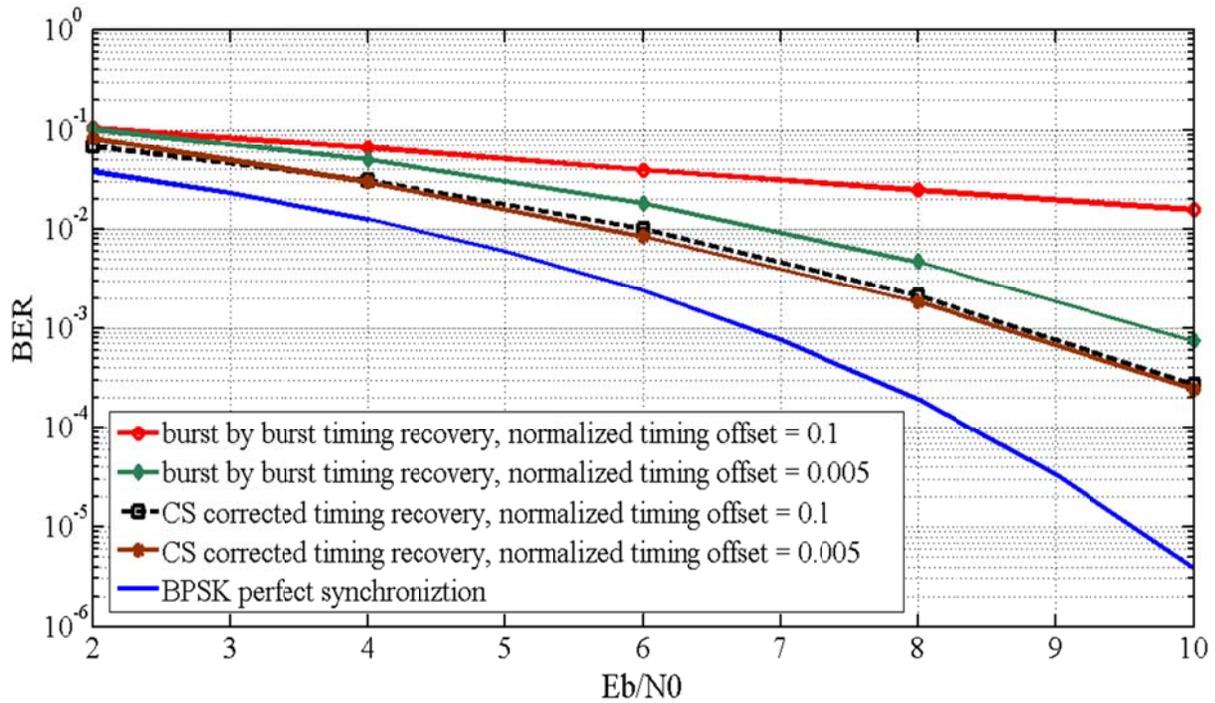

Fig. 7. typical approach and introduced algorithm BER performance for ϵ=0.1 and ϵ=0.005 , L=5000, N=300

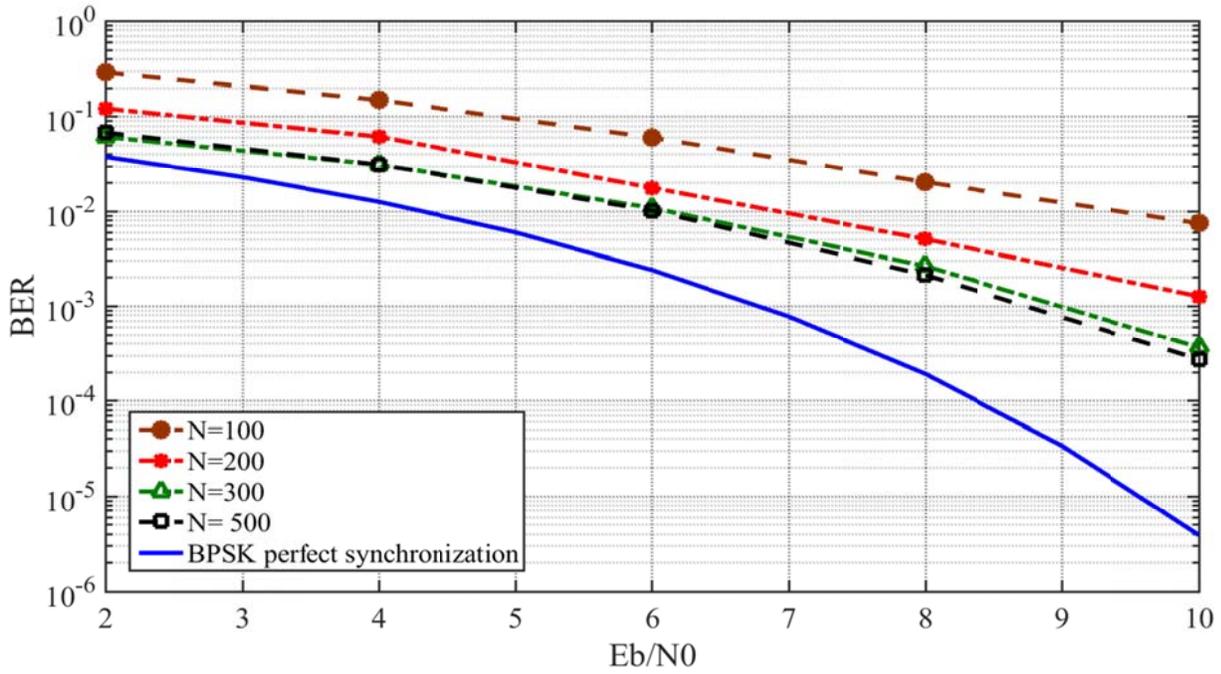

Fig. 8. BER performance for different burst lengths, $\epsilon = 0.1$, $L = 5000$

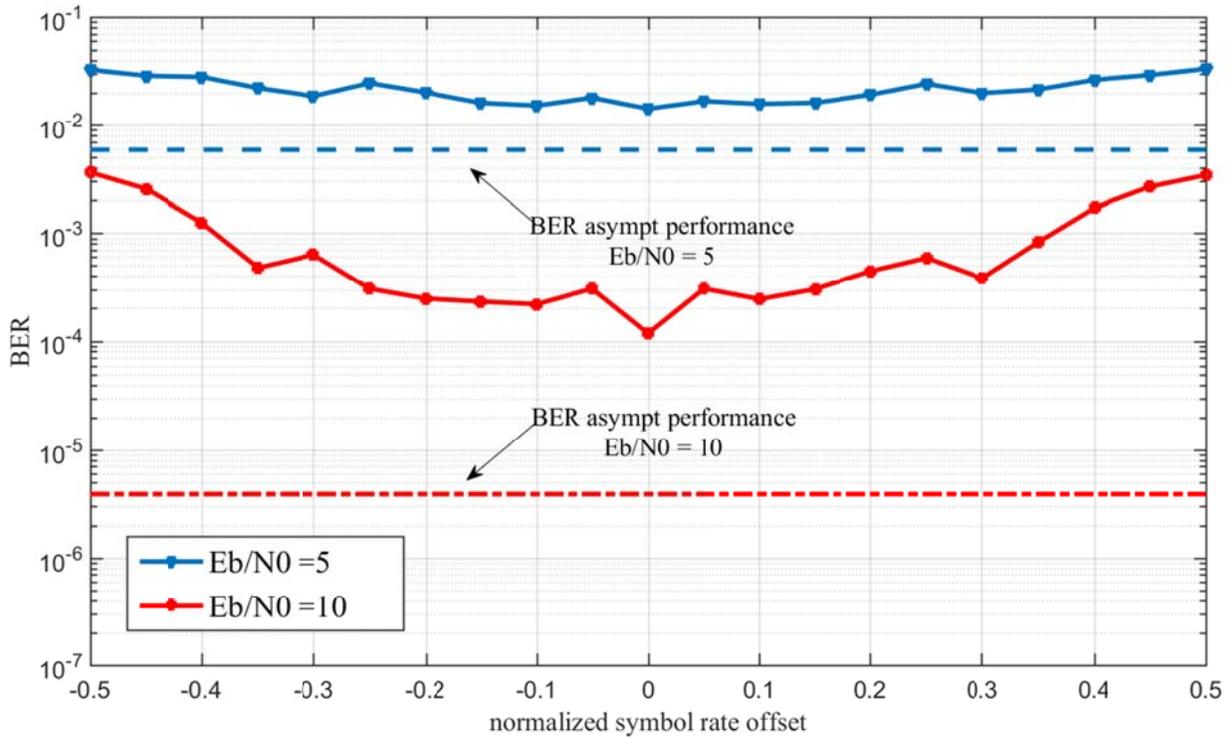

Fig. 9. BER performance versus normalized symbol rate offset, modulation type BPSK, burst length = 300, $L = 5000$

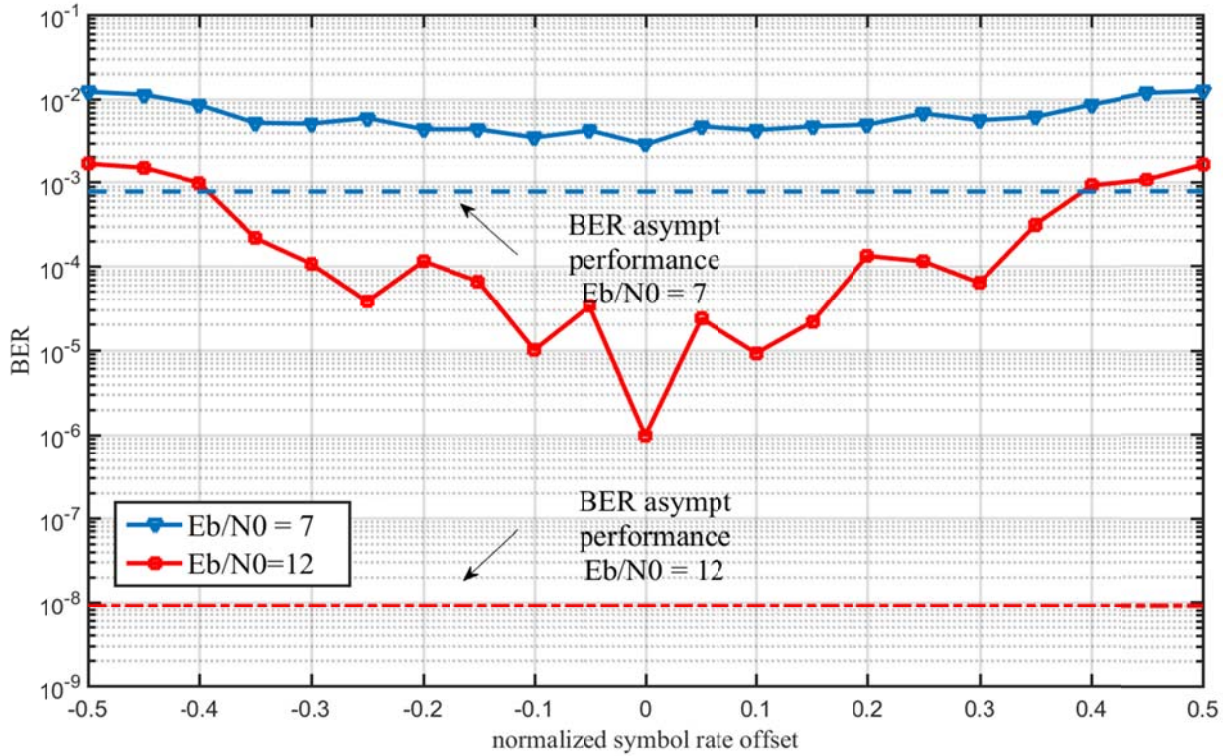

Fig.10. BER performance versus normalized symbol rate offset, modulation type QPSK, burst length = 300, $L = 5000$

It means that the received signal is divided into smaller bursts and independent timing recovery algorithm is applied to each segment. The length of each segment is calculated such that the timing error can be assumed to be fixed and thus CS happening is avoided in any segments. Clearly, for $\epsilon=0.1$ the shorter burst length will guarantee that CS does not take place than $\epsilon=0.005$, and as SNR increased the gap between typical approach and introduced algorithm widened and the poor performance eventuates for typical approach due to the short burst length. However, for the case that $\epsilon=0.005$, the performance of burst by burst algorithm is improved, due to the longer possible burst length. Meanwhile the proposed CS corrected algorithm achieves considerably better performance for both different $\epsilon$.

Figure.8 displays BER of BPSK signal for different burst lengths where normalized symbol rate offset $\epsilon$ is set to .1  As it is predictable, as the burst length increases the better result is achieved, since the longer bursts results in a better DFT resolution and more precise symbol rate modification in CS correction block.

In figure.9 and 10 the influence of different normalized symbol rate offsets on BER of BPSK and QPSK signal is depicted, where Eb/N0 assumed to be constant and burst length corresponds to 300 symbols. Overall declining trend in the performance is clearly demonstrated as the absolute value of normalized symbol rate offset increases, this is mainly due to this fact that larger offsets in symbol rate reduces the accuracy of timing error detected by TED. However, fluctuation of the plots' derivative, when system tends to achieve better performance despite increasing in $\epsilon$ (for example when $\epsilon$ corresponds to 0.3) can be interpreted through trade of between DFT resolution and TED's accuracy. More precisely $\epsilon$ augmentation leads to the faster CS happening and better DFT resolution due to the more periodic terms that is provided by larger $\epsilon$ in a fix burst length. While, in the other hand it reduces the accuracy of timing error detected by TED. Note that in CS inexistence ($\epsilon = 0$) system performance does not approach to the perfect synchronization, since system had to recover and estimate constant timing delay.

## 5 CONCOLUSION

A new algorithm for timing recovery by prior elimination of cycle slip is proposed. It is shown that linearly increased timing delay causes alternate cycle slips in synchronizer. A burst sequence of timing error provided by Gardner TED is used to indicate and eliminate CS. After CS correction iterative bandwidth efficient timing recovery is applied to the sequence of burst samples. The satisfactory simulation results, evaluated in terms of BER and compared with theoretical and other approaches confirm the system's appropriate performance.